\documentclass[pre,aps,onecolumn,superscriptaddress,notitlepage,nofootinbib]{revtex4-1}
\usepackage{amssymb,epsfig,amsmath,bm,subfigure,graphicx,textcomp,url,float,color,cases}

\usepackage{epsfig}
\usepackage{textcomp}
\usepackage[normalem]{ulem}

\usepackage[colorlinks=true, urlcolor=blue, anchorcolor=blue, citecolor=blue,filecolor=blue,linkcolor=blue,menucolor=blue
]{hyperref}

\usepackage{color}

\usepackage[titletoc,title]{appendix}

\begin{document}

\title{Full Statistics of Regularized Local Energy Density in a Freely Expanding Kipnis-Marchioro-Presutti Gas}
\author{Eldad Bettelheim}
\email{eldad.bettelheim@mail.huji.ac.il}
\affiliation{Racah Institute of Physics, Hebrew University of
Jerusalem, Jerusalem 91904, Israel}
\author{Baruch Meerson}
\email{meerson@mail.huji.ac.il}
\affiliation{Racah Institute of Physics, Hebrew University of
Jerusalem, Jerusalem 91904, Israel}
\begin{abstract}
We combine the Macroscopic Fluctuation Theory and the Inverse Scattering Method to determine the full long-time statistics of the energy density $u(x,t)$  averaged over a given spatial interval,
$$
U =\frac{1}{2L}\int_{-L}^{L}dx\, u(x,t)\,,
$$
in a freely expanding Kipnis-Marchioro-Presutti (KMP) lattice gas on the line, following the release at $t=0$ of a finite amount of energy at the origin. In particular, we show that, as time $t$ goes to infinity at fixed $L$, the large deviation function of $U$ approaches a universal, $L$-independent form when expressed in terms of the energy content of the interval $|x|<L$.  A key part of the solution is the determination of the most likely configuration of the energy density at time $t$, conditional on $U$.

\end{abstract}
\maketitle

\section{Introduction and MFT equations}
\label{intro}

Stochastic lattice gases  are simple, versatile and instructive models which
capture universal aspects of large deviations of different fluctuating quantities (the density field, the integrated current, \textit{etc.}) in  macroscopic systems of interacting particles out of equilibrium \cite{Spohn,Liggett,KL,Krapivskybook}. The last two decades have witnessed a major progress in this area of nonequilibrium statistical mechanics. To a large extent, this progress has been made possible by the development of fluctuational hydrodynamics (FHD) \cite{Spohn} and macroscopic fluctuation theory (MFT) \cite{JonaLasinioreview}, {the latter being a variant of the more general Optimal Fluctuation Method (OFM)}.  Stripped of irrelevant microscopic details, FHD and MFT are ideally suited for probing late-time asymptotic regimes in the lattice gases, which are usually the most interesting. This is especially true for non-stationary processes, where exact results for full statistics -- by any method -- are  scarce \cite{DG2009a,DG2009b,IMS2017,IMS2021,BSM2022a,
Grabsch2022,Mallick2022,BSM2022b,KLD2023,BM2024,Mallick2024}.

Here we focus on the continuous-time Kipnis-Marchioro-Presutti (KMP) lattice gas model \cite{KMP}. It involves immobile ``agents" occupying a lattice and carrying continuous non-zero amounts of ``energy". At each random move a randomly chosen pair of
nearest neighbors randomly redistributes their combined energy among them according to uniform distribution. The original motivation behind the KMP model was a rigorous proof of Fourier's law of heat diffusion when starting from a microscopic model \cite{KMP}. Since then the KMP model has been extensively studied in the context of nonequilibrium fluctuations of transport \cite{BSM2022b,BM2024,Bertini2005,BGL,BodineauDerrida,DG2009b,Lecomte,Tailleur,HurtadoGarrido,KrMe,
Pradosetal,MS2013,ZarfatyM,Peletier,Spielberg,Guttierez,Frassek,Mallick2024,Benichou}.

In a parallel development, the last two decades have witnessed remarkable advancement \cite{Corwin2012,Spohn2015,QuastelSpohn2015,Takeuchi2018} in the studies of the one-point statistics of the interface height at a specified point in space,
as described by the Kardar-Parisi-Zhang (KPZ) equation \cite{KPZ}, a paradigmatic model of nonconservative nonequilibrium stochastic interface growth. These works generated a renewed interest in the OFM as applied to the one-point statistics in the KPZ equation \cite{KK,MKV2016,KMS2016,JKM2016,SMP2017,SmithMeerson2018,KLD2021,KLD2022}.

A similar one-point statistics in a stochastic lattice gas would be the statistics of the gas density (or temperature) at a specified point in space.  It turns out, however, that in the continuum description of conservative lattice gases, provided  by the  FHD, the one-point statistics is ill-defined, as the variance of the one-point fluctuations diverges at small scales. In fact, a similar ultraviolet catastrophe of the one-point fluctuations in nonequilibrium stochastic growth is well known to experts \cite{Krug,SmithMS}. It occurs in a broad family of continuous stochastic interface growth models when the dimension of space exceeds a model-dependent critical dimension \cite{SmithMS}. Fortunately, for the KPZ equation the critical dimension is $2$,
which allows for unhindered studies of the one-point statistics in one dimension. For the diffusive lattice gases, however, the ultraviolet catastrophe occurs in all physical dimensions.

One possible regularization of the ultraviolet catastrophe would introduce a small-scale cutoff: most naturally, the lattice constant. Such a regularization, however,  would imply abandoning continuous theory with its many benefits. An advantageous regularization alternative is, therefore, to characterize local density/temperature fluctuations by the probability distribution $\mathcal{P}(U,T)$ of the density (or energy density) averaged over a small but \emph{macroscopic} spatial interval, $|x|<L$, at the observation time $t=T$:
\begin{equation}\label{BCT}
\frac{1}{2 L}\int_{-L}^{L} dx\,u(x,t=T) = U\,.
\end{equation}
This type of regularization was proposed in Ref. \cite{SmithMS} for the stochastic surface growth. Here we apply it to a freely expanding KMP gas with a finite total energy $W$. Since we are interested in the long-time limit, $T\to \infty$, we may assume that the energy is initially released at a point, which we  choose for simplicity to be $x=0$:
\begin{equation}\label{BC0}
u(x,t=0) = W \delta(x)\,.
\end{equation}
The \emph{expected} value of the locally averaged energy density $U$ at the same point at time $T$,
\begin{equation}\label{U0bar}
\bar{U} = \frac{W \text{erf}\left(\frac{L}{\sqrt{4T}}\right)}{2 L}\,,
\end{equation}
readily follows from the mean-field (that is, zero-noise) solution,
\begin{equation}\label{umf}
\bar{u}(x,t) = \frac{W}{\sqrt{4\pi t}}\, e^{-\frac{x^2}{4t}}\,,
\end{equation}
of the diffusion equation $\partial_t \bar{u} (x,t) = \partial_x^2 \bar{u}(x,t)$ that governs the mean coarse-grained energy density of the KMP gas at long times  \cite{KMP,Spohn,KL}. Here we study fluctuations, including large deviations, of $U$ around the expected value~(\ref{U0bar}). In the long-time limit, $T\gg L^2$,  the error function in Eq.~(\ref{U0bar}) can be expanded at small argument, and we obtain
\begin{equation}\label{U0barsmall}
\bar{U} \simeq  \frac{W}{\sqrt{4\pi T}}\,,
\end{equation}
which is independent of the regularization length $2L$.

It is convenient to rescale $t$, $x$ and $u$ by $T$, $\sqrt{T}$ and $W/\sqrt{T}$, respectively. The initial condition (\ref{BC0}) becomes
\begin{equation}\label{delta0}
u(x,t=0) = \delta(x)\,,
\end{equation}
while the condition (\ref{BCT}) on the locally averaged density at $t=T$ can be written as
\begin{equation}\label{density0}
\int_{-\ell}^{\ell} dx\,u(x,t=1)  = \kappa \equiv 2\ell \nu\,,
\end{equation}
where $\ell =L/\sqrt{T} \ll 1$. The quantity $\nu= U\sqrt{T}/W$ is the rescaled regularized local energy density, while $\kappa =2 \ell \nu$ is the fraction of the total energy observed at $t=T$ inside the interval $|x|<L$.

The (rescaled) optimal path of the process conditioned on $U$ is described by the well-known MFT equations for the KMP model \cite{JonaLasinioreview,DG2009b}. They can be recast in the following symmetric form \cite{MS2013,BSM2022a}:
\begin{eqnarray}
  \partial_t u &=& \partial_x (\partial_x u+2 u^2 v)\,, \label{d1} \\
  \partial_t v &=& \partial_x (-\partial_{x} v+2 u v^2)\,. \label{d2}
\end{eqnarray}
Equations~(\ref{d1}) and (\ref{d2}) follow from a minimization of the action functional of a proper path integral, constrained by Eq.~(\ref{density0}), with respect to variations of $u(x,t)$ \cite{JonaLasinioreview,DG2009b,MS2013,BSM2022a}. The minimization also yields, aside from Eqs.~(\ref{d1}) and~(\ref{d2}), a boundary condition at the (rescaled) final time,
\begin{equation}
\label{vdelta}
v(x,t=1) = \lambda\left[\delta(x-\ell)-\delta(x+\ell)\right]\,,
\end{equation}
where $\lambda$ is a Lagrange multiplier, to be ultimately expressed through $\nu$ and $\ell$ with the help of Eq.~(\ref{density0}) \cite{SmithMS}. Importantly, the MFT equations and boundary conditions for this problem obey the mirror symmetry and antisymmetry relations
\begin{equation}\label{symmetry}
u(-x,t) = u(x,t)\quad \text{and} \quad v(-x,t) = -v(x,t)\,.
\end{equation}

Once $u(x,t)$ and $v(x,t)$ have been found, one can calculate the rescaled action $s$ \cite{DG2009b,MS2013}
\begin{eqnarray}
  s  = \int_0^1 dt \int_{-\infty}^{\infty} dx \, u^2(x,t) \,v^2(x,t) = 2\int_0^1 dt \int_{0}^{\infty} dx \, u^2(x,t) \,v^2(x,t)
  \,.\label{action0}
\end{eqnarray}
It is often simpler, however, to determine the action by using the ``shortcut relation"
\begin{equation}\label{shortcutrelation}
\frac{ds}{d\kappa}=\lambda(\kappa)\,,
\end{equation}
which follows from $\kappa$ and $\lambda$ being conjugate variables, see \textit{e.g.} Ref. \cite{shortcut}. The action gives the desired probability distribution $\mathcal{P}$ up to a sub-leading, pre-exponential factor that we do not attempt to evaluate:
\begin{equation}\label{scalingc}
\ln {\mathcal P}(U,T,\ell)\simeq
-\sqrt{T} \,s\left(\nu,\ell\right).
\end{equation}
Since $\sqrt{T} \gg 1$, Eq.~(\ref{scalingc}) clearly has a large-deviation structure, with the rescaled action $s\left(\nu,\ell\right)$ playing the role of a rate function.

The main objective of this work is to determine the large-deviation function $s\left(\nu,\ell\right)$ in the leading order in the small regularization parameter $\ell$ but for  arbitrary $\nu$. In Sec. \ref{lineartheory} we describe typical, small fluctuations of $\nu$ around its expected value $\bar{\nu}=\bar{U} \sqrt{T}/W$. Here one can use the Lagrange multiplier $\lambda$ as a small expansion parameter \cite{KrMe}. We also present in Sec. \ref{lineartheory} a free-energy argument that, as we argue, applies in a broad range of the densities $\nu$. In Sec. \ref{ISM} we apply the Inverse Scattering Method to obtain important relations which, {for sufficiently small $\ell$,}  enable us to determine the rate function $s\left(\nu,\ell\right)$, to justify the free-energy argument and establish its applicability limits, and to find additional asymptotics of the rate function.  In particular, we show that, in the limit of  $\ell\to 0$ at constant $\kappa=2\ell \nu$, the rate function expressed in terms of $\kappa=2\ell \nu$ acquires a universal ($\ell$-independent) form. We also present two instructive examples of the optimal configurations $u(x,t=1)$ at small $\ell$ and different $\lambda$ (or $\nu$), as predicted by the ISM, as well as an example of the complete optimal path, computed by solving the MFT equations numerically.  In Sec. \ref{discussion} we briefly summarize and discuss our results.

\section{Linear theory and free-energy argument}
\label{lineartheory}

Typical fluctuations of $\nu$ around its expected value $\bar{\nu}=\bar{U} \sqrt{T}/W$ are Gaussian. Here one can linearize Eqs.~(\ref{d1}) and (\ref{d2}) with respect to $|\lambda|\ll 1$ \cite{KrMe}.
Then Eq.~(\ref{d2}) becomes simply $\partial_t v = -\partial_x^2 v$. This anti-diffusion equation can be solved backward in time with the ``initial" condition~(\ref{vdelta}). The solution, for arbitrary $\ell$, is
\begin{equation}\label{vlinear}
v(x,t) = \frac{\lambda}{\sqrt{4\pi(1-t)}}\,\left[e^{-\frac{(x-\ell)^2}{4 (1-t)}}-e^{-\frac{(x+\ell)^2}{4
   (1-t)}}\right]\,.
\end{equation}
The action $s$ in the leading (second) order in $\lambda$ can be obtained from Eq.~(\ref{action0}) by using the solution~(\ref{vlinear}) for $v(x,t)$ and setting $u(x,t)=\bar{u}(x,t)$ from (a rescaled) Eq.~(\ref{umf}). Evaluating the double integral, we arrive at the following expression:
\begin{equation}\label{slinlambda}
s(\lambda,\ell)=\frac{\lambda ^2} {4 \sqrt{2 \pi }}
  \left[e^{-\frac{\ell^2}{2}}-\text{erfc}\left(\frac{\ell}{\sqrt{2}}\right)
  \right]\,.
\end{equation}
Then, using the shortcut relation (\ref{shortcutrelation}), we can express the rescaled action in terms of $\nu$:
\begin{equation}\label{slinnu}
s(\nu,\ell)=\frac{4 \sqrt{2 \pi } \,\ell^2\, e^{\frac{\ell2}{2}}}{1-e^{\frac{\ell^2}{2}} \text{erfc}\left(\frac{\ell}{\sqrt{2}}\right)}\, \left(\nu-\bar{\nu}\right)^2\,,
\end{equation}
where
\begin{equation}\label{nubar}
\bar{\nu} \equiv \frac{\bar{U}\sqrt{T}}{W} = \frac{1}{2\ell}\,\text{erf}\left(\frac{\ell}{2}\right)\,,
\end{equation}
and the validity of the linear theory requires that $|\nu-\bar{\nu}| \ll \bar{\nu}$.

For $\ell \ll 1$, that is at sufficiently long time at fixed $L$, Eqs.~(\ref{slinnu}) and (\ref{nubar}) simplify to
\begin{equation}\label{linearssmalll}
s(\nu,\ell)\simeq 4 \pi \ell \left(\nu-\bar{\nu}\right)^2
\end{equation}
and
\begin{equation}\label{nubarsmall}
\bar{\nu}\simeq \frac{1}{\sqrt{4 \pi}}\,,
\end{equation}
respectively. The variance of fluctuations of $\nu$ is equal to $1/(8\pi \ell)$. It diverges in the limit of $\ell \to 0$, exhibiting the ultraviolet catastrophe \cite{SmithMS}. The $1/\ell$ scaling of the variance is to be expected. Indeed, the final rescaled energy density inside the small interval  $|x|<\ell$ is already almost constant and equal to $1/(8\pi \ell)$. In its turn, the variance of $\nu$ scales as one over the energy content of this interval.

For future reference, the small-$\ell$ asymptotic of the action ~(\ref{slinlambda}) in terms of $\lambda$ is
\begin{equation}\label{slinlambdasmalll}
s(\lambda,\ell\ll 1)=\frac{\ell \,\lambda^2}{4\pi}\,.
\end{equation}
Alongside with the shortcut relation~(\ref{shortcutrelation}), this again leads to the small-$\ell$ asymptotic (\ref{linearssmalll}).

When $\ell$ is small,  the linear theory predicts that the optimal density profile at the observation time in the ``interior region" $|x|\ll \ell$ is  flat:
\begin{equation}\label{u01}
  u(|x|<\ell,t=1) = \frac{1}{\sqrt{4\pi}}+\frac{\lambda}{4\pi}+\dots \,,
\end{equation}
where the ellipsis denotes sub-leading corrections in $\ell\ll 1$.
By assuming a flat density profile $u(|x|<\ell,1)=\nu$  from the start, one can obtain Eq.~(\ref{u01}) from Eqs.~(\ref{shortcutrelation}), (\ref{linearssmalll}) and (\ref{nubarsmall}).

Importantly, at  $\ell \ll 1$, the gas in the small interior region $|x|<\ell$ has a sufficient time to reach thermal equilibrium \cite{SmithMS}. It is not surprising, therefore,  that the Gaussian asymptotic~(\ref{slinlambdasmalll}), and hence (\ref{linearssmalll}), for the action also follows from the Boltzmann-Gibbs free energy difference \cite{DG2009b,KrMe}
\begin{equation}
\label{freenergy}
  s \simeq \int_{-\ell}^{\ell} dx \left[F(u(x,1))-F(\bar{u})-F^{\prime}(\bar{u})(u(x,1)-\bar{u})\right]\,,
\end{equation}
where $F(u) = -\ln u$ is the equilibrium free energy density of the KMP gas, see e.g. Refs. \cite{DG2009b,KrMe}. To see it, one should  set in Eq.~(\ref{freenergy}) $\bar{u} =\bar{\nu}= 1/\sqrt{4 \pi}$, use Eq.~(\ref{u01}) for $u(|x|<\ell,1)$ and expand the integrand in powers of $\lambda$ up to, and including, a quadratic term.

As we will show in Sec. \ref{ratefunction}, at $\ell \ll 1$ Eq.~(\ref{freenergy}) actually holds well beyond the linear theory, once the density at the final time $t=1$ is approximately uniform in the interior region: $u(|x|<\ell,1) \simeq \nu= \text{const}$. In this case Eq.~(\ref{freenergy}) predicts
\begin{equation}
\label{freenergyresult}
  s_{\text{equil}}(\nu,\ell) \simeq 2\ell \left(\frac{\nu}{\bar{\nu}}-\ln\frac{\nu}{
  \bar{\nu}}-1\right)\,.
\end{equation}
Notice that the equilibrium rate function~(\ref{freenergyresult}) diverges logarithmically at $\nu \to 0$, {implying zero probability of forming a void -- that is, a zero-energy region of a finite size -- in the KMP model \cite{KMSvoid}}.  This important property is characteristic of the KMP gas, and it can be traced down to the nature of its microscopic model, which deals with a continuous energy variable rather than with discrete particles.

\section{Solution by the Inverse Scattering Method}
\label{ISM}

\subsection{Solving the scattering problem: general}
\label{scattering}

It was noticed in Ref. \cite{BSM2022a} that the MFT equations (\ref{d1}) and (\ref{d2}) formally coincide with the derivative nonlinear Schr\"{o}dinger equation (DNLSE) in imaginary space and time  \cite{Shabat}. Application of the Inverse Scattering Method (ISM) to the DNLSE was pioneered by Kaup and Newell \cite{KaupNewell} for initial-value problems. Adaptation of the method to boundary-value problems in time has been described in detail in several papers \cite{BSM2022a,BSM2022b,KLD2023,BM2024}. Nevertheless, we outline  here the main steps of the method, as it applies to the current problem, forgoing a derivation of Eq. (\ref{TEvolution}) below, and skipping some of the actual calculations, as they can be easily recovered. Following Kaup and Newell \cite{KaupNewell}, one starts with the auxiliary scattering problem for the {vector} field $\boldsymbol\psi(x,t;k)$,
\begin{align}
&\partial_x \boldsymbol{\psi}(x,t;k)=V(x,t;k)\boldsymbol{\psi}(x,t;k)\,,\label{inversescatt}
\end{align}
where
\begin{align}
&V(x,t;k)=\begin{pmatrix}
-\frac{\zeta^2}{2} & -i  v(x,t)\zeta \\
-i  u(x,t)\zeta &  \frac{\zeta^2}{2} \\
\end{pmatrix}
\end{align}
and $\zeta=\sqrt{i k}$. The variable $k$ is a  wave number  describing the plane-wave boundary conditions at $x\to\infty$ for $\boldsymbol{\psi}(x,t;k)$, while the $t$-dependence comes into play in the form of dependence of $V$  on $t$ through $u$ and $v$. We omit here those details of the scattering theory which we do not use directly, such as the time dependence of
$\psi$, referring the reader to Ref. \cite{BSM2022a} where more details are given. At this point,  one assumes $\boldsymbol{\psi}(x,t;k)=\begin{pmatrix}\alpha e^{-i k x} \\
\beta e^{i k x} \\
\end{pmatrix}$ at $x\to-\infty$. Then, using the linearity {of Eq.~(\ref{inversescatt})} and the asymptotic behavior of the solution at infinity, where $u$ and $v$ vanish, one obtains  the following asymptotic behavior of $\boldsymbol{\psi}(x,t;k)$ at $x\to+\infty$:
\begin{align}
\boldsymbol{\psi}(x,t;k)=\begin{pmatrix}\alpha a(k;t) e^{-i k x}+\beta \tilde b(k;t) e^{-i k x} \\
\alpha b(k;t)e^{i k x}+\beta \tilde a(k;t)e^{i k x} \\
\end{pmatrix}\,.
\end{align}
The integrability of the problem manifests itself in the fact that the time evolution of the parameters $a,$ $\tilde a$, $b$ and $\tilde b$ is very simple:
\begin{align}
\label{TEvolution}a(k;t)=a(k;0), \quad \tilde a(k;t)=\tilde a(k,0),\quad e^{k^2t}b(k,t)=b(k,0), \quad\text{and}\quad \tilde b(k,t)=\tilde b(k,0)e^{k^2t}\,.
\end{align}
These relations, alongside with the fact that $a(k;t)$ is analytic in the upper half $k$-plane, while $\tilde{a}(k;t)$ is analytic in the lower half plane \cite{KaupNewell}, allows one to find explicit expressions for $v(x,0)$ and $u(x,1)$, as we will see below.
Another useful relation, although we will not explicitly use it here, is given by $a(k;t) \tilde a(k;t)-b(k;t)\tilde b(k;t)=1$.

Now we proceed to calculating $\tilde{b}(k;0)$  and $ \tilde{a}(k;0)$ by solving the scattering problem, Eq. (\ref{inversescatt}),  at time $t=0$ setting $\alpha=0$ and $\beta=1$.   The solution of the scattering problem is greatly aided by the fact that $u(x,0)=\delta(x)$, see Eq.~(\ref{delta0}).  We obtain
\begin{align}
 \tilde{b}(k;0)=\frac{\tilde{a}(k;0)a(k;0)-1}{-i \zeta}, \quad b(k;0)=-i\zeta, \quad \quad \tilde{a}(k;0)=1-i k Q_-(k)\,\quad {a}(k;0)=1-i k Q_+(k),\label{0time}
\end{align}
where
\begin{align}
Q_+(k)=\int_0^\infty e^{ik x}v(x,0)dx, \quad Q_-(k)=\int_{-\infty}^0 e^{ikx}v(x,0)dx \quad \text{and}\quad Q(k)=Q_+(k)+Q_-(k)\,. \label{Qdef}
\end{align}

A similar exercise can be performed  at $t=1$, this time making use of the simple form of $v(x,1)$, given by Eq.~(\ref{vdelta}),
to solve the scattering problem, Eq. (\ref{inversescatt}). Let us define the following quantity, which will prove useful in the sequel:
\begin{align}
I_a^b(k)=\int_{a}^b e^{-i k x}u(x,1)dx\,.
\end{align}
This produces the result
\begin{align}
& \tilde{b}(k;1)= \lambda^2k  \zeta I^{+\ell}_{-\ell}+2   \lambda \zeta\sin(k\ell), \quad \tilde a(k,1)=1+2\lambda kI_{\ell}^\infty \sin(k\ell)+\lambda i k e^{-i k \ell}I^{+\ell}_{-\ell}+\lambda^2 k^2I^{+\ell}_{-\ell}I^\infty_{\ell}.\label{Fullab}\end{align}
One could also derive equations for the other scattering parameters, $a$ and $b$, at $t=1$, and then use Eq. (\ref{TEvolution}) to obtain a set of closed nonlinear integral equations for $u$ and $v$, as it was recently done in another context in Ref. \cite{GRB2024}. 
It is not known, however, how to solve these nonlinear integral equations analytically, and their numerical solution offers no clear advantage over directly solving the original MFT equations (\ref{d1}) and (\ref{d2}).

We thus omit the expressions for  $a$ and $b$  for the sake of brevity and proceed by invoking a simplifying asymptotic limit of small $\ell$ {which, as explained in the Introduction, is directly relevant in the context of the regularized local energy density at long times}.  In the next subsection, we use this limit to obtain closed-form expressions for $u(x,1)$ and $v(x,0)$, up to a single parameter which can be computed numerically. This parameter is the value of the rescaled regularized local energy  $\kappa $,  and the equation to be solved numerically is  Eq.~(\ref{integralEq}) below.

\subsection{{Solving the scattering problem for small $\ell$}}
\label{smallell}

For small $\ell$ it is natural to assume that the integral $I_{-\ell}^\ell(k)$ is approximately independent of $k$. Indeed,  $I_{-\ell}^\ell(k)$ is the Fourier transform of a function, $u(x,1)H(\ell-|x|)$ (where $H(x)$ is the Heaviside function), which is concentrated in a vanishingly small region around the origin. Essentially,  this function  behaves like a delta function, therefore its Fourier transform is nearly constant. We proceed with this assumption and check its validity \textit{a posteriori}.

{Assuming that} $I_{-\ell}^{\ell}(k)$ is a constant, we can write  $I_{-\ell}^{\ell}(k)\simeq I_{-\ell}^{\ell}(0)\equiv\kappa $ [see Eq. (\ref{density0})]. Now, instead of having to determine the function $I_{-\ell}^\ell(k)$ in order to solve the problem, we only have to determine self-consistently the parameter $\kappa.$  In the limit of small $\ell,$ Eq. (\ref{Fullab}) yields the following approximations for $\tilde b$ and $\tilde a$:
\begin{align}
&\tilde b (k;1)\simeq-i\lambda\zeta^3(\lambda \kappa+2\ell)\,,\label{1}\\
&\label{2}\tilde{a}(k;1)\simeq1+ \lambda i k \kappa+\lambda k^2(\lambda \kappa+2\ell)\, I_\ell^\infty.
\end{align}

We now make use of the simple relation between $\tilde b(k;1)$ and $\tilde b(k;0)$, given by Eq. (\ref{TEvolution})  and the equation for $\tilde b(k;0)$, given in Eq. (\ref{0time}), together with the expression in Eq. (\ref{1}), to write an equation for $Q_\pm$:
\begin{align}
&\label{keyeq}\left(1-i kQ_+(k)\right)\left(1-i kQ_-(k)\right)=1+\lambda(\lambda \kappa+2\ell)k^2e^{-k^2}\,.
\end{align}
Similar equations have recently appeared in the ISM as applied to several MFT and OFM settings \cite{KLD2023,BSM2022a,Mallick2022,BSM2022b,KLD2022,KLD2023,BM2024,Mallick2024}.
Equation~(\ref{keyeq}) can be solved by the Wiener-Hopf method. In our case, however, this equation still has one undetermined parameter $\kappa$. To obtain an equation for this parameter we can use the simple relation between  $\tilde a(k;1)$ and $\tilde a(k;0)$, given in Eq. (\ref{TEvolution}), and the equation for $\tilde a(k;0)$, given in Eq. (\ref{0time}) together with the expression in Eq. (\ref{2}). We obtain
\begin{align}
\lambda\kappa- i k\lambda(2\ell+\kappa\lambda )\int_{\ell}^{\infty} e^{-i k x}u(x,1)dx=-  \int_{-\infty}^0 e^{i k x}v(x,0)dx.\label{uvrelation}
\end{align}
Integrating the second term by parts and making an obvious change of variables in the last term, we can rewrite this equation in the following form:
\begin{align}
\lambda\kappa-\lambda (2\ell+\kappa\lambda )e^{-i k \ell}u(\ell^+,1)=\int_{\ell}^{\infty} \lambda (2\ell+\kappa\lambda )e^{-i k x}u'(x,1)dx + \int_{0}^\infty e^{-i k x}v(x,0)dx,
\end{align}
where $u(\ell^+,1)$ denotes the limiting value of $u(x,1)$ as $x$ approaches the point $x=\ell$ from the right, and we have used the antisymmetry relation $v(-x,t)=-v(x,t)$.  {For} small $\ell$ one then obtains, {to  leading order in $\ell$}, the following relation:
\begin{align}
\lambda\kappa-\lambda (2\ell+\kappa\lambda )u(\ell^+,1)=\int_{\ell}^{\infty} e^{-i k x}\left[\lambda (2\ell+\kappa\lambda )u'(x,1)\,\,+v(x,0)\right]dx, \label{key}
\end{align}
Sending $k$ to $\infty$, we observe that the integral on the right hand vanishes as $1/k$, since $u'(x,1)$ and $v(x,0)$ are regular at $x=\ell$.  Therefore, the left hand side, being independent  of $k$,  must vanish, and we obtain
\begin{equation}\label{u0uplus}
\kappa=(2\ell+\lambda\kappa)\,u(\ell^+,1).
\end{equation}
Then, since the left hand side of Eq. (\ref{key}) is zero for any $k$, the right hand side must vanish identically, which leads to
\begin{align}
 &v(x,0)=-\lambda(2\ell+\lambda\kappa)  u'(x,1).\label{vuprimerelation}
\end{align}
Combining Eq. (\ref{u0uplus}) and (\ref{vuprimerelation})
we obtain the desired equation for $\kappa$:
\begin{equation}\label{u0isvint}
  \kappa= \frac{ 1}{\lambda}  \int_{0}^{\infty} v(x,0)dx=\frac{ Q_+(0)}{\lambda}  \,\,.
\end{equation}

Together, Eqs. (\ref{keyeq}) and (\ref{u0isvint}) give a closed set of equations for $v(x,0)$, since $Q_\pm(k)$ are the Fourier transforms of  $v(x,0)H(\pm x)$. Indeed,  as already mentioned, the  quantities $1-i kQ_\pm$ are analytic in the upper and lower half $k$-planes, respectively. Therefore, we can interpret this equation as the decomposition of the function on the right hand side into a product of functions analytic in the upper and lower half planes, respectively. It is then natural to take the logarithm of this equation. This is facilitated by the following definition of $M_\pm$:
\begin{align}
1-i kQ_\pm(k)=e^{M_\pm}.\label{RiemannHopf}
\end{align}
The decomposition of the logarithm of the right hand side of Eq. (\ref{keyeq}) into $M_\pm$,  which in turn is the logarithm  of   $1-i k Q_\pm,$ is performed in the usual way to give the following result:
\begin{align}
&M_\pm=\pm\int_{-\infty}^{\infty} \frac{\ln\left(1+\lambda\left(2\ell+  \lambda \kappa\right) k'^2e^{-k'^2}\right)}{k'-k\mp i0^+} \frac{dk'}{2\pi i },
\label{q+q-firstEq.}
\end{align}
which yields an expression for $Q_\pm$:
\begin{align}
\label{basically the result}i  k Q_\pm(k)=1-\exp\left[\pm\int_{-\infty}^{\infty} \frac{\ln\left(1+\lambda\left(2\ell+  \lambda \kappa\right) k'^2e^{-k'^2}\right)}{k'-k\mp i0^+} \frac{dk'}{2\pi i }\right].
\end{align}

We note that since $u(x,1)$ vanishes at infinity, the functions  $Q_\pm(k)$ are regular at the origin, which means that the left hand side vanishes as $k\to0$. One then verifies that the right hand side vanishes in this limit as well. This check is necessary, because an additive constant may, in principle, appear on the right hand side of Eq. ({\ref{q+q-firstEq.}), a possibility that we have tacitly ignored above.

It is now possible to find $Q_+(0)$ from Eq. (\ref{basically the result})
and plug it into Eq. (\ref{u0isvint}). This calculation yields a closed equation for $\kappa$ (or $\nu=\kappa/2\ell$),
\begin{align}
\lambda \kappa=\int_{-\infty}^{\infty} \frac{\ln\left[1+\lambda\left(2\ell+  \lambda \kappa\right) k^2e^{-k^2}\right]}{k^2} \frac{dk}{2\pi}\,,
\label{integralEq}
\end{align}
which can be solved numerically.

We note here an important point about the choice of branch of the logarithm in Eqs. (\ref{basically the result}) and (\ref{integralEq}). A linear analysis shows that for small $|\lambda|$ (the corresponding strong inequality for $\lambda$ depends on $\ell$) the correct solution is obtained by choosing the standard branch of the logarithm. One can ask then whether a branch point of the logarithmic function can cross the real $k$-axis at larger $|\lambda|$. If this happens, localized soliton-like solutions may appear \cite{KLD2022,BSM2022b}, and a careful analysis of the choice of the proper branch of the logarithmic function must be performed. Specifically for the KMP model, the localized solution -- coupled propagating $u$- and $v$ pulses -- is known as a doublon \cite{MS2013,ZarfatyM,BSM2022b}. Importantly, in the regime of small $\ell$ that we are dealing with here, the doublons can be ruled out on the physical grounds \cite{MS2013,ZarfatyM}.

At this point we also mention that the assumptions that we have made about the solution at small $\ell$, \textit{e.g.} that  $I_{-\ell}^\ell(k)$ is independent of $k$, that $\sin(k\ell)$ can be replaced by $k\ell$, \textit{etc.}, can now be confirmed. This is done by examining relations (\ref{TEvolution}), which now have explicit expressions associated with them, and taking the small $\ell$ limit. The analysis is aided by the fact that integrals, such as those appearing in (\ref{q+q-firstEq.}) have dominant contributions in the region $k\ll\ell^{-1}$.

Equation~(\ref{integralEq}), combined with the shortcut relation (\ref{shortcutrelation}), suffices for the purpose of calculating the action $s(\nu,\ell)$ which serves as the rate function of the probability density $\mathcal{P}(U,T,\ell)$, see Eq.~(\ref{scalingc}). In their turn, Eqs.~(\ref{basically the result}) and~(\ref{integralEq}) yield the optimal solution for the conjugate field $v(x,0)$, while
the relationship~(\ref{vuprimerelation}) between $v(x,0)$ and $u'(x,1)$ allows one to find the optimal energy density $u(x,1)$ as well.

\subsection{Rate function and its asymptotics}
\label{ratefunction}

For general $\lambda,$ Eq.~(\ref{integralEq}) for $\nu$ can only be solved numerically. As an example, the left panel of Fig.~\ref{nuvslambda} shows a fragment of the resulting dependence of $\nu$ upon $\lambda$ for $\ell=0.01$. The point $\lambda=0$ corresponds to the mean-field value $\nu=\bar{\nu}=1/\sqrt{4\pi}$, as to be expected from Eq.~(\ref{nubarsmall}).  The right panel of Fig.~\ref{nuvslambda}  shows the rate function $s(\nu)$ obtained from  Eq.~(\ref{shortcutrelation}) by integrating numerically the inverse function $\lambda(\nu)$. The inset compares this $s(\nu)$ with predictions from the linear theory, Eq.~(\ref{linearssmalll}), and  from the free-energy difference, Eq.
(\ref{freenergyresult}). As one can see, the Gaussian asymptotic, predicted by the linear theory, applies only in a narrow region of $\nu$ around $\bar{\nu}$, whereas the free-energy difference captures $s(\nu)$ in a much broader range of $\nu$.

\begin{figure}[ht]
\includegraphics[width=0.4\textwidth,clip=]{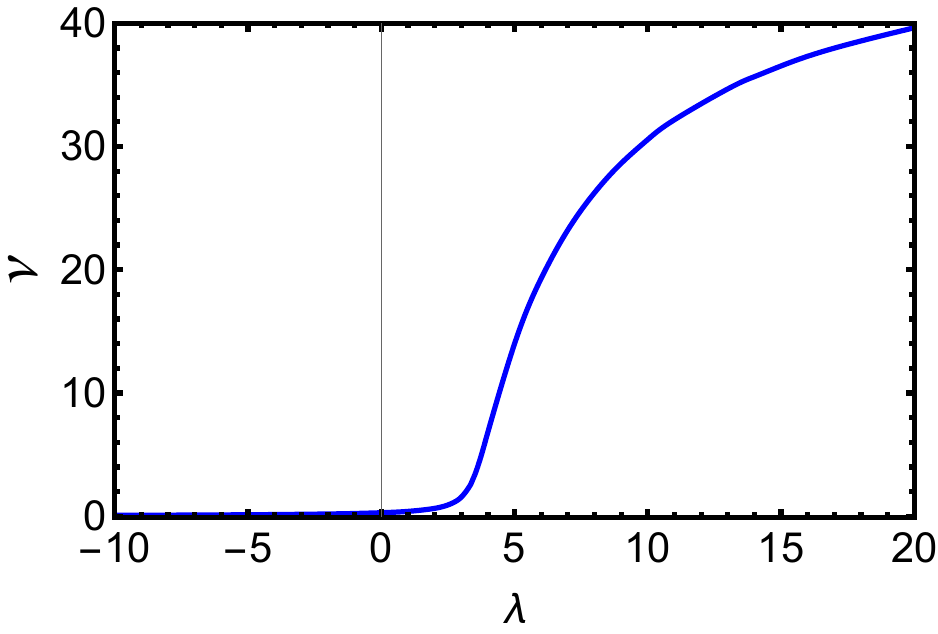}
\includegraphics[width=0.4\textwidth,clip=]{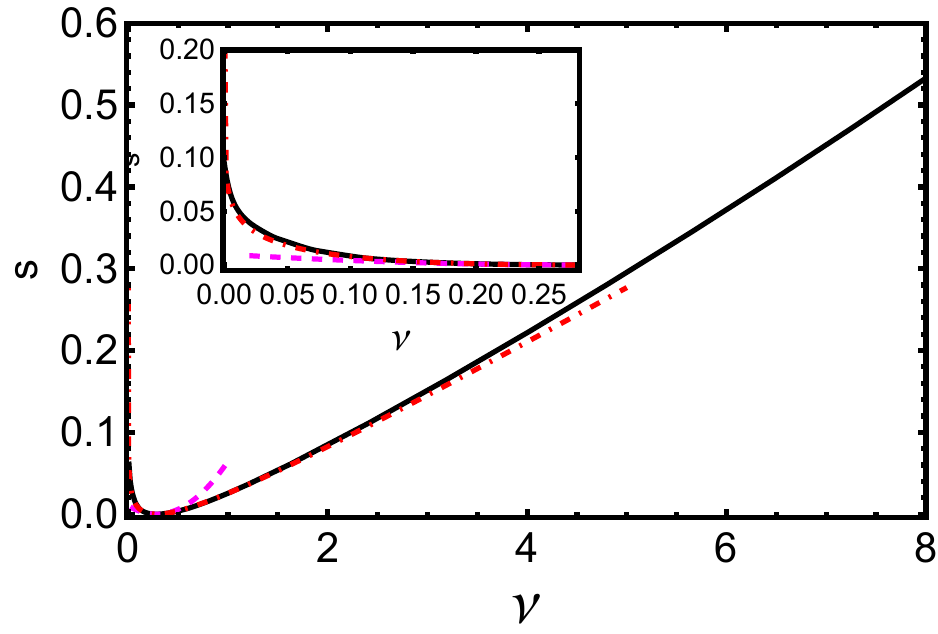}
\caption{Left panel: $\nu=\nu(\lambda)$, found by numerically solving Eq.~(\ref{integralEq}) for $\ell=0.01$ and different $\lambda$. Right panel:  the rate function $s$ vs. $\nu$ for this $\ell$ (the solid line), the linear theory prediction~(\ref{linearssmalll}) (the dashed line) and the prediction from the free-energy difference (\ref{freenergyresult}) (the dot-dashed line). The inset shows a blowup of the interval $0<\nu<\bar{\nu}$.}
\label{nuvslambda}
\end{figure}

To understand the formal reason behind the success of the free-energy prediction, we notice that, because of the factor $e^{-k^2}$ of the integrand in Eq.~(\ref{integralEq}), the characteristic width of the integration region in $k$ is $O(1)$. Let us assume (and check \textit{a posteriori}) that
\begin{equation}\label{fe0}
2|\lambda\ell \left(1+\lambda \nu\right)|\ll 1\,,
\end{equation}
and expand the logarithm keeping only the linear term: $\ln\left[1+2\lambda\ell \left(1+\lambda \nu\right) k^2e^{-k^2}\right]\simeq 2\lambda\ell \left(1+\lambda \nu\right) k^2e^{-k^2}$. Evaluating the elementary integral and solving the resulting simple algebraic equation for $\nu$, we obtain
\begin{equation}\label{fe1}
\nu = \frac{1}{\lambda_c-\lambda}\,,\quad\text{where} \quad \lambda_c=\sqrt{4\pi}\equiv 1/\bar{\nu}\,.
\end{equation}
Now integrating the shortcut relation~(\ref{shortcutrelation}) with $\kappa = 2\ell \nu = 2\ell/(\lambda_c-\lambda)$, taken from Eq.~(\ref{fe1}), we reproduce the free-energy prediction~(\ref{freenergyresult}). As we already discussed, the  Gaussian asymptotic (\ref{linearssmalll}), predicted by linear theory, is  a particular limit of Eq.~(\ref{freenergyresult}), only valid in the vicinity of $\nu=\bar{\nu}$. To obtain this asymptotic directly  from Eq.~(\ref{integralEq}), one should expand the integrand in the powers of $\lambda$ up to and including the second order, and keep only the leading (first-order) term in $\ell\ll 1$.

Importantly, for Eq.~(\ref{fe1}) to make sense, $\lambda$ must obey the condition
\begin{equation}\label{conditionleft}
-\infty <\lambda<\lambda_c\,.
\end{equation}
In addition, the assumed strong inequality (\ref{fe0}) can be justified only if $\lambda$ is not too close to $\lambda_c$: $\lambda_c -\lambda \gg \ell$. As a result, the free-energy prediction (\ref{freenergyresult}) for the rate function applies for all $\nu$ much smaller than $O(1/\ell)\gg 1$.

What happens at $\nu\gtrsim O(1/\ell)$, {that is far from equilibrium}? Here the rate function can be determined numerically from Eqs.~(\ref{shortcutrelation}) and~(\ref{integralEq}), as shown in Fig. \ref{nuvslambda}. However, an important simplification occurs in the limit of $\ell \to 0$ while keeping $\kappa = 2\nu \ell$ constant.  As can be checked \textit{a posteriori}, here the term $\lambda \nu$ inside the parentheses in Eq.~(\ref{integralEq}), is much larger than $1$. Neglecting this $1$ in the zeroth order in $\ell$ at fixed $\kappa = 2\ell \nu$, one arrives at a universal  equation for $\kappa = 2\ell \nu$ versus $\lambda$:
\begin{equation}
2 \pi\lambda \kappa=\int_{-\infty}^{\infty} \frac{\ln\left(1+\lambda^2 \kappa k^2e^{-k^2}\right)}{k^2} dk\,.
\label{universaleq}
\end{equation}
The resulting function $\kappa(\lambda)$, obtained numerically, is shown on the left panel of Fig. \ref{kappauniversal}. In this limit $\kappa(\lambda)$ vanishes identically for all $\lambda<\lambda_c$, and it undergoes a tangent bifurcation at $\lambda=\lambda_c$. The right panel of Fig. \ref{kappauniversal} shows the resulting universal rate function $s_>(\kappa)$ alongside with its two asymptotics: $s_>(\kappa\to 0) \simeq \sqrt{4\pi}\, \kappa$,  and the asymptotic at $\kappa\to 1$. Importantly, the $\kappa \to 0$ asymptotic matches the $\nu\gg \bar{\nu}$ asymptotic of the equilibrium rate function, which is described by the first term in the r.h.s. of Eq.~(\ref{freenergyresult}).

The asymptotic at $\kappa\to 1$ can be expressed in terms of product log (Lambert $W$) function, very similarly to how it has been done in Ref. \cite{BSM2022a}.  Alternatively, it can be presented in a parametric form as follows:
\begin{equation}\label{largekappa}
  \kappa(\lambda\to \infty) \simeq 1-\frac{6\sqrt{\ln \lambda}}{\pi\lambda}\,,\quad
  s_>(\lambda\to \infty) \simeq \frac{4 (\ln\lambda)^{3/2}}{\pi }\,.
\end{equation}

\begin{figure}[ht]
\includegraphics[width=0.4\textwidth,clip=]{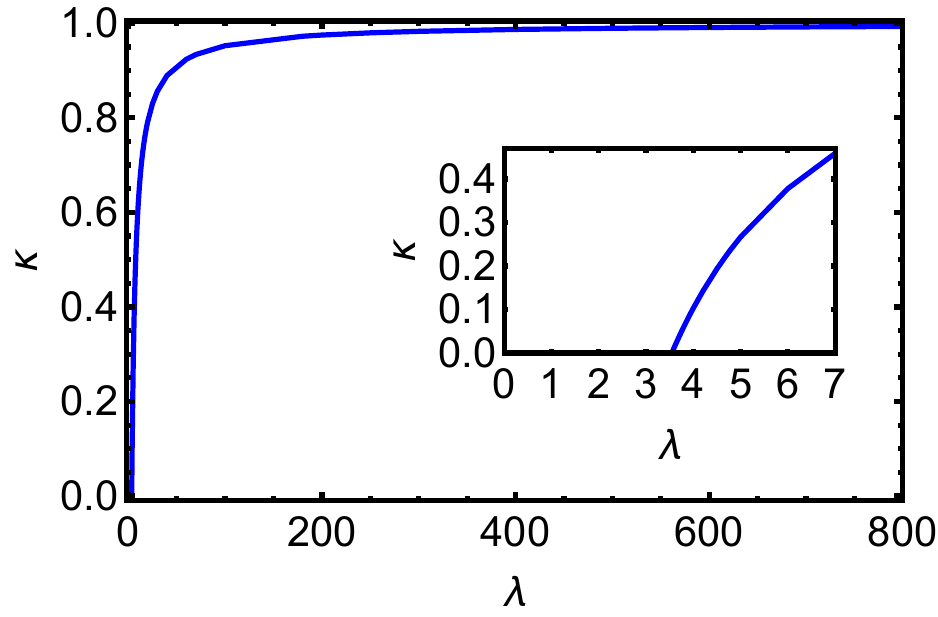}
\includegraphics[width=0.4\textwidth,clip=]{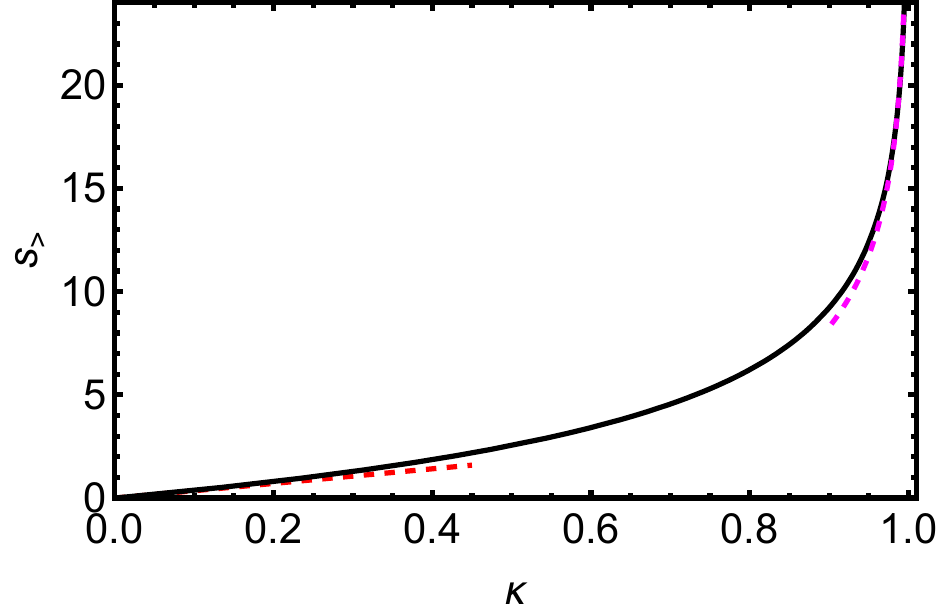}
\caption{Left panel: $\kappa=2\ell\nu$ versus $\lambda$, found by numerically solving the universal ($\ell \to 0$) equation (\ref{universaleq}). The inset focuses on the region close to the tangent bifurcation point $\lambda=\lambda_c= \sqrt{4\pi}$. Right panel:  the rate function $s_>(\kappa)$ in terms of the energy fraction $\kappa$ alongside with its small-$\kappa$ asymptotic $s_>(\kappa\to 0) \simeq \sqrt{4\pi}\, \kappa$, and the large-$\kappa$ asymptotic~(\ref{largekappa}).}
\label{kappauniversal}
\end{figure}

When $\ell$ is small but finite, the tangent bifurcation at $\lambda=\lambda_c$ is smoothed out, and $\kappa$ is non-zero even for $\lambda<\lambda_c$. As $\ell$ goes down, the functions $\kappa(\lambda,\ell)$, calculated from Eq.~(\ref{integralEq}) for different $\ell$, converge to the universal curve
$\kappa(\lambda)$. This convergence in evident in Fig.~\ref{convergenceell}: at $\ell=10^{-4}$ the function $\kappa(\lambda,\ell)$ is already almost indistinguishable from the universal curve shown in the inset of the left panel of Fig. \ref{kappauniversal}.

\begin{figure}[ht]
\includegraphics[width=0.4\textwidth,clip=]{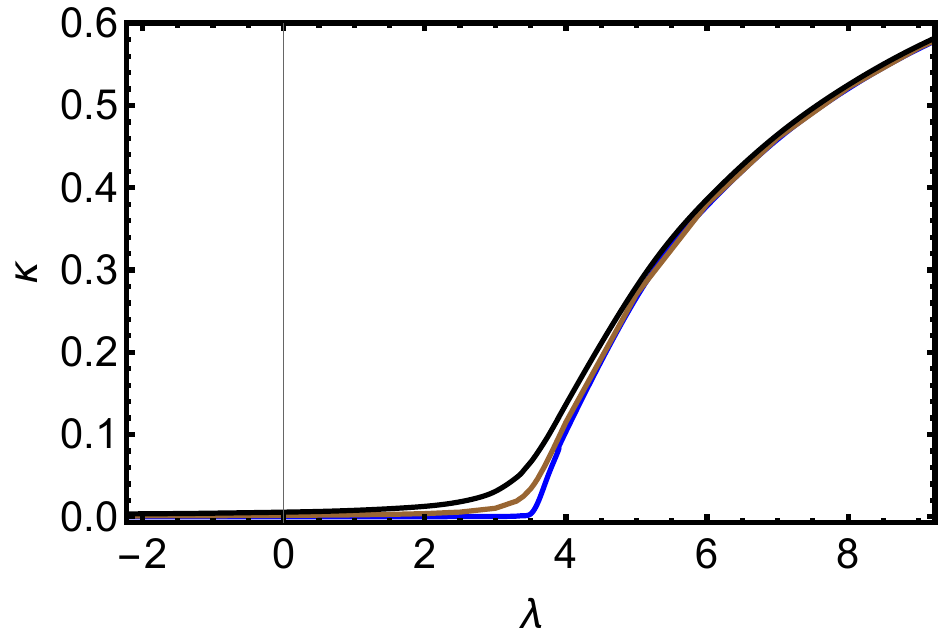}
\caption{Convergence of $\kappa(\lambda,\ell)$ to the universal function $\kappa(\lambda)$ for $\ell=10^{-2}$, $3\cdot 10^{-3}$ and $10^{-4}$ (from top to bottom).}
\label{convergenceell}
\end{figure}

{Now we are in a position to examine the validity of our analytical results.
The main assumption we made in order to arrive at Eqs.~(\ref{1}) and (\ref{2}) is that, to a leading order in $\ell$, the integral $\int_{-\ell}^\ell e^{-i k x }u(x,1)$ is  independent of $k$ in the relevant range of $k$ in the problem, and therefore it can be approximated by $2\nu\ell$. This requires that the characteristic length scale of $u(x,1)$, which is $O(\ell)$, be much smaller than the relevant $1/k$. The latter is determined by the effective width of the integration region in Eq.~(\ref{integralEq}). For $-\infty<\lambda<\lambda_c$, $\nu$ is given by Eq.~(\ref{fe1}). As one can check, in this case the relevant range of $k$ is $O(1)$, and the strong inequality $\ell\ll 1$ suffices.}

For $\lambda\gtrsim \lambda_c$ an additional condition appears. This is because in this case we obtain $\nu\ell \simeq 1$. As a result, the effective width of the integration region over $k$ in Eq.~(\ref{integralEq}) is $O(\sqrt{\ln \lambda^2})$. This quantity should be much smaller than $1/\ell$:
\begin{equation}\label{logcondition}
\ell \sqrt{\ln \lambda^2} \ll 1\,.
\end{equation}
This condition invalidates the rate function, that we found here,
when $\kappa$ becomes too close to its (unreachable) maximum possible value $1$.

\subsection{{Optimal paths}}
\label{optimalpaths}

\begin{figure}[ht]
\includegraphics[width=0.39\textwidth,clip=]{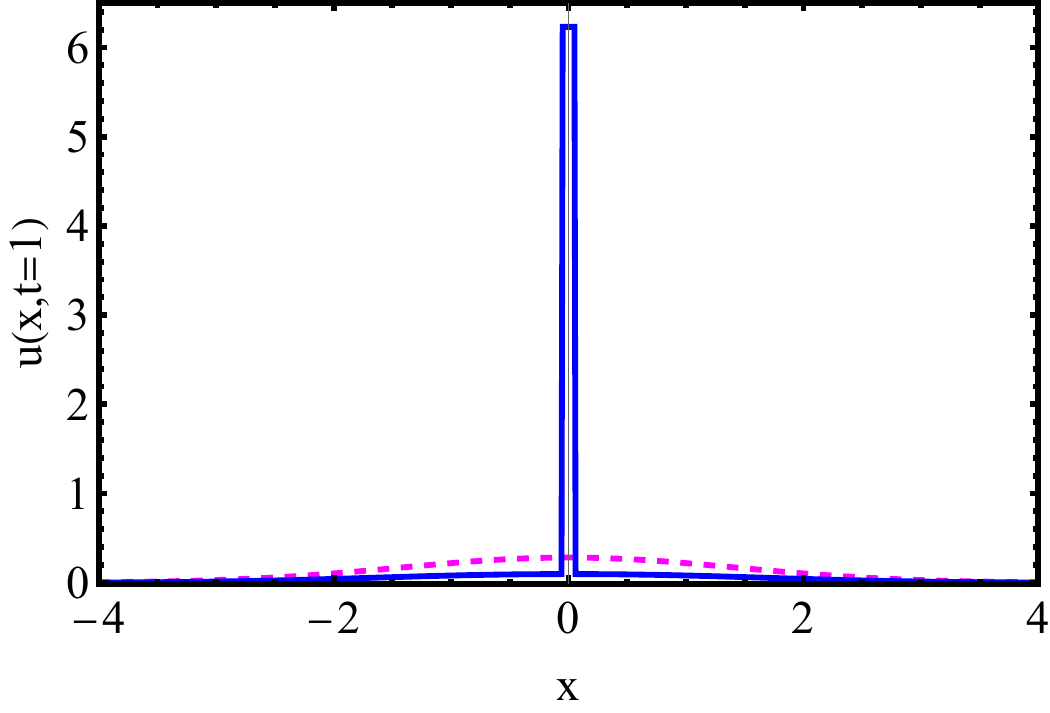}
\includegraphics[width=0.4\textwidth,clip=]{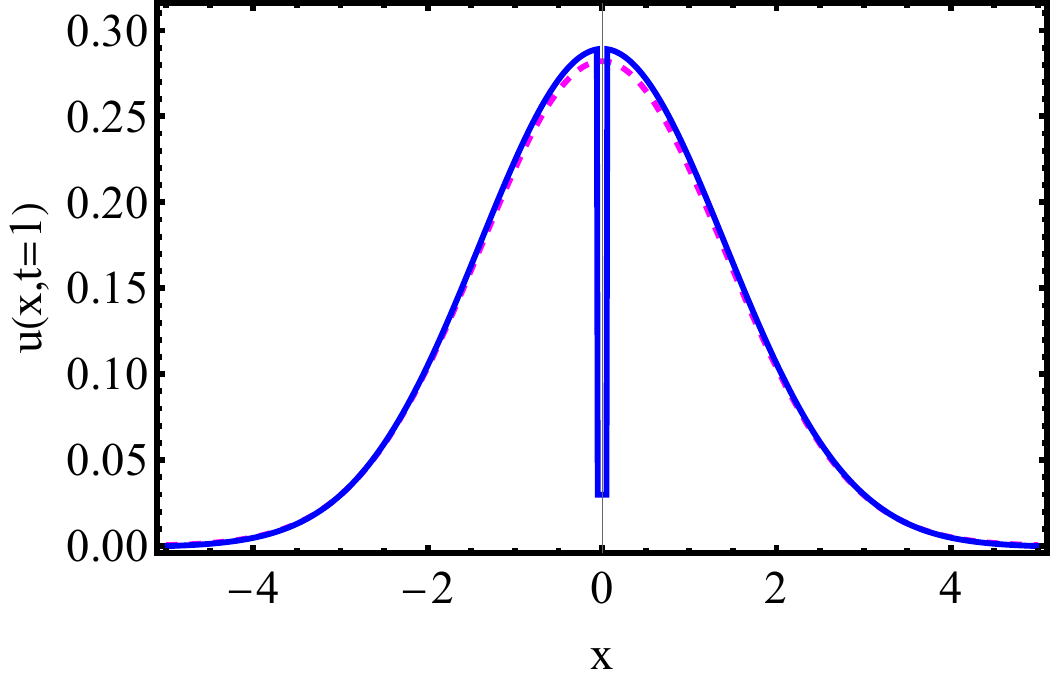}
\caption{Solid lines: $u(x,1)$ for $\lambda =10$ (left panel)   and $-30$ (right panel) as predicted by the ISM for $\ell=0.05$. Dashed lines: the expected  (mean-field) density profile $(4\pi)^{-1/2}e^{-x^2/4}$. Note the different vertical scales.}
\label{ux1}
\end{figure}

As we have seen, the ISM also predicts the optimal density profile $u(x,1)$ and the optimal profile of the conjugate field $v(x,0)$. These can be found numerically, in a straightforward manner, from Eqs. (\ref{vuprimerelation}), (\ref{basically the result}) and (\ref{integralEq}). Two examples of the optimal density profiles $u(x,1)$ for $\lambda=10$ and $\lambda=-30$ are shown in  Fig. \ref{ux1}.  For comparison, also shown is the expected, that is mean-field, density profile $e^{-x^2/4}/\sqrt{4\pi}$. {As one can see, the two optimal density profiles in these cases are very different and quite instructive.}

{By virtue of the shortcut relation (\ref{shortcutrelation}) one does not have to know the full optimal path $u(x,t)$ and $v(x,t)$ for the calculation of the rate function $s\left(\nu,\ell\right)$. It is still interesting, however, to compute the optimal path, as it provides a valuable insight into the mechanism of the large deviation in question.  Here the ISM, which focuses on $u$ and $v$ only at $t=0$ and $t=1$, is insufficient, and one has to resort to a full numerical solution of the nonlinear partial differential equations (\ref{d1}) and (\ref{d2}) with the boundary conditions (\ref{delta0}) and (\ref{vdelta}). Fortunately,  such a solution -- by the back-and-forth iteration algorithm due to Chernykh and Stepanov \cite{CS} is available and has by now become standard. As an example, Fig. \ref{ell1} shows some snapshots of the optimal path for $\ell =1$ and $\lambda=5$, that we computed with this algorithm. One salient feature of this finite-$\ell$ setting (in contrast to the small-$\ell$ limit we dealt with in most of Sec. \ref{ISM}) is a visibly non-uniform density profile in the internal region $|x|<\ell$.}

\begin{figure}[ht]
\includegraphics[width=0.40\textwidth,clip=]{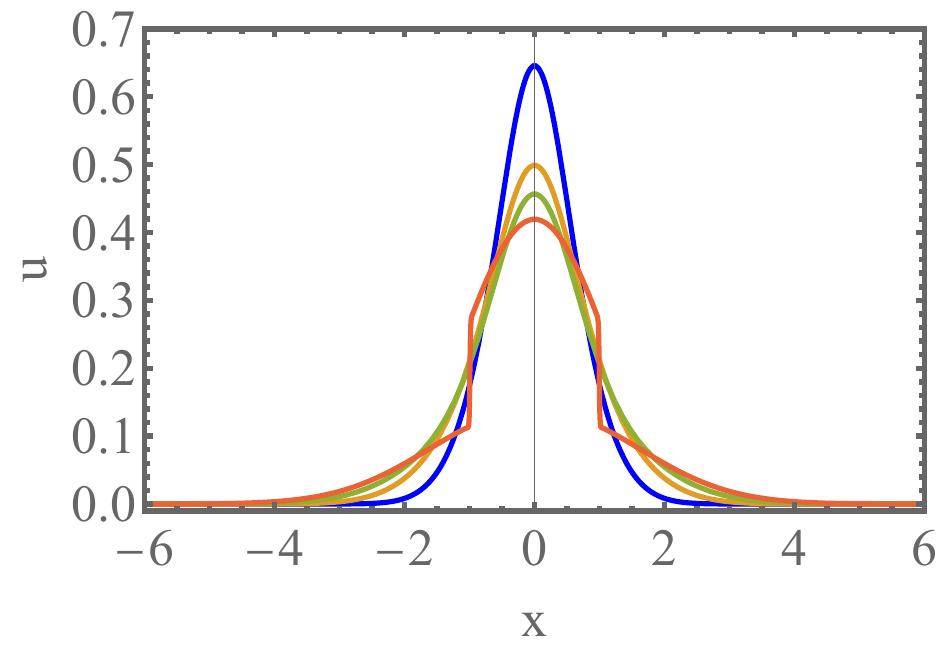}
\includegraphics[width=0.40\textwidth,clip=]{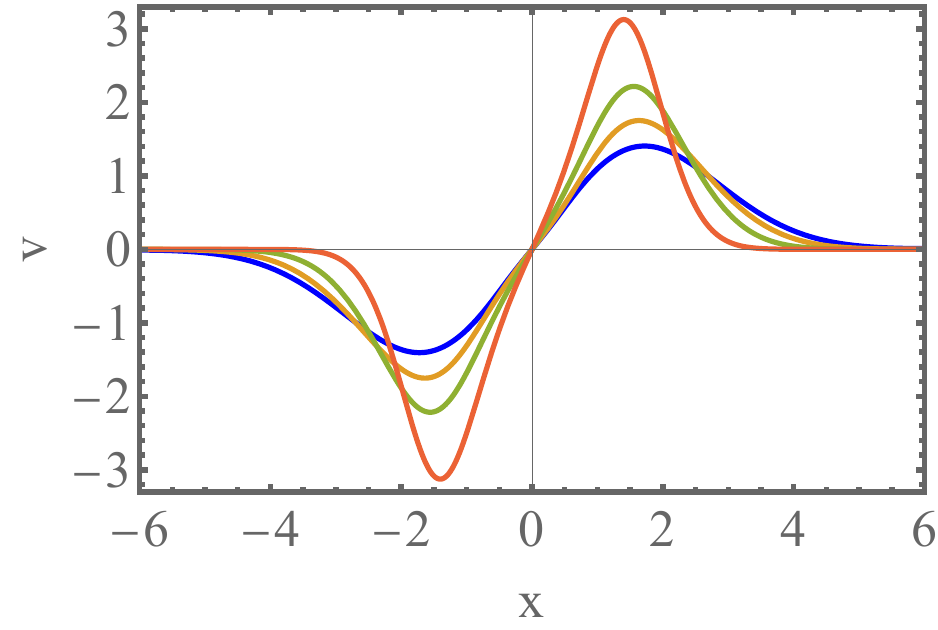}
\caption{{The optimal path for  $\ell=1$ and $\lambda=5$, computed with the back-and-forth iteration algorithm \cite{CS}. Shown are  $u(x,t)$ at $t=1/4$, $1/2$, $3/4$ and $1$ (left panel)   and $v(x,t)$ at $t=0$, $1/4$, $1/2$ and $3/4$ (right panel).}}
\label{ell1}
\end{figure}

\section{Summary and Discussion}
\label{discussion}
By combining the MFT and the ISM, we determined the long-time large-deviation statistics of the locally-averaged energy density in a freely expanding Kipnis-Marchioro-Presutti lattice gas. We showed that the corresponding rate function exhibits two distinct regions. One region corresponds to relatively small local densities and is described by the equilibrium free-energy argument. The other region, associated with large densities, is fundamentally non-equilibrium. As time increases, the rate function as a function of the energy content of the interval $|x|<\ell$ converges to a universal, $\ell$-independent form.

An attempt to solve this problem without relying on the strong inequality $\ell\ll 1$ would be an interesting endeavor. Such an extension is not required in the context of regularization of the single-point density. On the other hand, it would make a natural connection to the statistics of accumulated current of matter or energy in a specified region of space \cite{DG2009a,DG2009b}. There is also an interesting formal analogy between the setting with arbitrary $\ell$ and the problem of finding the joint distribution of two accumulated currents in the Simple Symmetric Exclusion Process.  The latter problem has been recently formulated in terms of the ISM  (for the nonlinear Schr\"{o}dinger equation rather than for the DNLSE),  and a possible approach to the solution has been explored \cite{GRB2024}.

At $\ell\gtrsim 1$ and sufficiently small $\nu$ (that is, sufficiently large negative $\lambda$), one should expect a propagating doublon to appear, which will affect the choice of the branch of the logarithm in Eqs. (\ref{basically the result}) and (\ref{integralEq}). The physical mechanism behind the doublon formation is quite simple \cite{BSM2022a,MS2013,ZarfatyM}: it pays off in terms of the action for the fluctuations to form a propagating doublon if the process is conditioned on evacuation of a sufficiently large amount of energy from a sufficiently long interval $|x|<\ell$. This regime is quite interesting, as one would expect a breaking of mirror symmetry (\ref{symmetry}) in this case. The reason is that the cost, in terms of the action, of two mirror-symmetric outgoing doublons is expected to be higher than that of a single doublon with a larger amplitude that transports the same amount of energy \cite{ZarfatyM}.

It is worth mentioning that the governing MFT equations~(\ref{d1}) and (\ref{d2}) can be also obtained for all diffusive lattice gases with a constant diffusivity and a quadratic density dependence of the mobility. However, the delta-function initial condition (\ref{delta0}), which was crucial in our solution, can be relevant only for those lattice gases where the local density is unbounded from above. These include, for example, the Simple Symmetric Inclusion Process (SIP) \cite{BM2024,GRV2010}, but exclude the Simple Symmetric Exclusion Process (SSEP) \cite{Spohn,Liggett,KL}.

In conclusion, over the past three years, the ISM has been instrumental in solving an increasing number of field-theoretical large-deviation problems \cite{KLD2021,KLD2022,BSM2022a,Mallick2022,BSM2022b,KLD2023,BM2024,Mallick2024,KLD2024}. The present work, however, appears to be the first instance where the ISM has been combined with a perturbation approach exploiting a small parameter (in this case the rescaled regularization length $\ell$) unrelated to a Lagrange multiplier.  We believe that this approach holds significant promise.

\subsection*{Acknowledgments}
We are very grateful to N. R. Smith for a critical reading of the paper and insightful comments. We also thank A. Grabsch and O. B\'{e}nichou for attracting our attention to
their recent work \cite{GRB2024}. This research was supported by the US-Israel
Binational Science Foundation through Grant No. 2020193 (E.B.) and by the Israel
Science Foundation through Grant 1499/20 (B.M.).


\end{document}